\documentclass[showpacs]{revtex4}
\usepackage{amsmath}
\usepackage{amssymb}
\usepackage{graphicx}

\begin{document}

\title{Nonrelativistic isothermal fluid in the presence of a chameleon scalar field:\\
 Static and collapsing configurations
 }

\author{
Vladimir Folomeev
\footnote{Email: vfolomeev@mail.ru}
}
\affiliation{
Institute of Physicotechnical Problems and Material Science of the NAS
of the
Kyrgyz Republic, 265 a, Chui Street, Bishkek, 720071,  Kyrgyz Republic \\
}
\begin{abstract}
We consider a gravitating spherically symmetric nonrelativistic configuration consisting of
a  massless chameleon scalar field nonminimally coupled to
a perfect isothermal fluid.
The object of this paper is to show the influence of the chameleon scalar field on the structure and evolution of an  isothermal sphere.
For this system we find  static, singular and regular solutions depending on the
form of the coupling function.
A preliminary stability analysis indicates that both stable and unstable solutions exist.
For unstable configurations, by choosing
the special form of the coupling function,
we consider the problem of the gravitational collapse by applying the similarity method.
\end{abstract}

\maketitle

\section{Introduction}

Over  the  course  of  the  past  decades,
there has been substantial progress in cosmology and astrophysics supported
by a growing quantity and quality of observational data  requiring its theoretical description.
Difficulties associated with such description
are obvious: in considering various objects and processes in the Universe,
one must deal with scales which differ by  orders of magnitude.
It forces the use of only approximate models and methods permitting   considerable simplification
in describing the structure and evolution of the Universe.
In this respect, various scalar fields are ones of the most called-for objects which are being used in investigations of the Universe
on all scales. They are  relevant in describing   the early inflationary Universe \cite{Linde},  the present accelerated
Universe  \cite{Copeland:2006wr},
and the processes on scales of order of galaxies and separate stars.
In particular,
this has come to refer to attempts of describing the dark matter in galaxies \cite{Bertone:2004pz} and to a consideration of compact configurations --
boson stars supported by various scalar fields \cite{Schunck:2003kk}.

In modeling boson stars, it  usually assumes that they consist of complex or real scalar fields in their own gravitational field.
However, it is also possible to imagine a situation when besides a scalar field there can exist other forms of matter in a system.
This can be fermion fields \cite{Schunck:2003kk}, an electromagnetic field
\cite{Matos:1995my}, or ordinary matter. In the latter case, ordinary matter can interact with a scalar field either only via gravity or
by use of   nonminimal coupling as well. In the present paper we consider a case
of the presence of such nonminimal coupling in a compact gravitating configuration.
As applied to cosmology, the idea of  nonminimal coupling between a scalar field
and ordinary matter in the form of a perfect fluid was suggested in Refs.~\cite{Khoury:2003aq,Khoury:2003rn,Brax:2004qh},
where the nonminimal coupling was used in describing the accelerated expansion of the present Universe.
In this case the effective mass of the scalar field can change  depending on
 the background environment. Because of it, such a scalar field was called a ``chameleon scalar field.''
 The idea that the properties of a scalar field could be influenced
by the environment/matter surrounding the scalar field was studied earlier,
in particular,
 in the papers \cite{Ellis:1989as,Mota:2003tc}
where the interaction between matter and the scalar field was used to model a dependence of
fundamental  coupling  constants on the local environment.
Further development of this idea can be found in the papers
\cite{Farajollahi:2010pk, Cannata:2010qd}, where the authors
  describe different forms of cosmological evolution within the framework of the chameleon cosmology.

In progressing from cosmological to astrophysical scales,
in a recent work \cite{arXiv:1106.1267} we considered  the model
 of chameleon stars consisting of a scalar field nonminimally coupled to ordinary matter
in the form of a perfect polytropic fluid having the equation of state $p \sim \rho^\gamma$,
where $\rho$ and $p$ are the density and the pressure of the fluid,
and $\gamma$ is some constant. For this system we found  static, regular, asymptotically flat solutions for both relativistic and
nonrelativistic cases for $\gamma=2$ and $5/3$. For these values of $\gamma$,
it was shown that the presence of the nonminimal interaction leads to substantial changes
both in the radial matter distribution of the star and in the star's total mass.

In the present paper we continue studying compact astrophysical objects consisting of a scalar field nonminimally
coupled to  the polytropic fluid having $\gamma=1$. This case corresponds to the isothermal gas sphere whose temperature is assumed to
be constant along the radius. In the nonrelativistic case, there are known static
singular and regular solutions describing  such equilibrium configurations
in weak gravitational fields \cite{Chandra:1939}. Depending on the parameters of the model,
they can be gravitationally stable or unstable. In the latter case, instability leads to motion of matter that can be realized
in the form of collapse.
 The gravitational
collapse of an isothermal gas cloud is a well-studied problem.
In different variants, this problem has been investigated  frequently.
In particular, there is a number of works devoted to the consideration of self-similar  motions
 of an isothermal gas. We indicate some of them here.

In considering the collapse, there has been a great deal of attention  paid to the specification  of initial conditions.
Larson \cite{Larson:1969mx} and Penston \cite{Penston:1969yy} chose
zero initial velocity and uniform density of the fluid. Such conditions provide
 a homologous inflow that evolves to a central region. This solution assumes that
 there exists the flow at large radii, or equivalently as $t\to 0$,
which is directed inward at $3.3$ times the sound speed  and the density in the envelope is proportional to $r^{-2}$.
It gives the density be $4.4$ times the  value appropriate to hydrostatic equilibrium.
Shu \cite{Shu:1977uc} criticizes this solution
and suggests new self-similar solutions, which did not have these problems.  He assumes the
 initial density profile in the form $\rho \sim r^{-2}$. Such a sphere is quasistatic,
so that the infall velocities are negligible at the moment of protostar formation.
The resulting initial configuration
is the singular isothermal sphere, which is an unstable hydrostatic
equilibrium.
Shu showed that in this case
 the cloud collapse begins at the center, and the infall spreads outward. He called
this solution the expansion-wave solution.
Hunter \cite{Hunter:1977}  has been continuing the work in that direction, and
found a new class of similarity solutions which includes
  Shu's expansion-wave solution as one limit. He considered different initially unstable spheres
and studied whether their collapses are well described by similarity solutions.

In this paper we consider the model of a nonrelativistic configuration consisting of a perfect
isothermal fluid in the presence of a  massless chameleon scalar field.
Our intention here is to clarify the influence of the
 chameleon scalar field on such a widely studied  system as the gravitating isothermal fluid.
The paper is organized as follows: In Sec.~\ref{static_eqs_rel} we derive the equations
for a static configuration with the arbitrary  coupling function $f$ (see the Lagrangian \eqref{lagran_cham_star_col} below).
Using these equations,  in next three subsections  we consider various static configurations for different
choices of $f$. For such configurations, we find analytical and numerical solutions, and also
address the issue of their stability. Using as initial conditions  unstable static singular spheres from
Sec.~\ref{choice_f_exp}, in Sec.~\ref{collaps_sing}  
we consider similarity solutions describing the collapse of such configurations.
Our conclusions are summarized in
the final section, where we also suggest some possible lines of further  investigation.

\section{Static configurations}
\label{static_eqs_rel}

 As discussed in the Introduction we consider a gravitating system of a real scalar field coupled to a perfect fluid.
In the general case of strong gravitational fields,
 the Lagrangian for this system is
\begin{equation}
\label{lagran_cham_star_col}
L=-\frac{c^4}{16\pi G}R+\frac{1}{2}\partial_{\mu}\varphi\partial^{\mu}\varphi -V(\varphi)+f(\varphi) L_m ~.
\end{equation}
Here $\varphi$ is the real scalar field with the potential $V(\varphi)$; $L_m$ is the Lagrangian of the perfect
isotropic fluid i.e. a fluid with only one radial pressure; $f(\varphi)$ is some function describing the
nonminimal interaction between the fluid and the scalar field. An interaction similar to this was used
to describe the evolution of dark energy within the framework of chameleon cosmologies \cite{Farajollahi:2010pk, Cannata:2010qd}.
The case $f=1$ corresponds to the absence of the nonminimal coupling. In this case the two sources
are still coupled via gravity.

The above Lagrangian was used by us in
\cite{arXiv:1106.1267} to construct static relativistic and nonrelativistic configurations. In that case,
the Lagrangian for the isentropic  perfect fluid was chosen
to have the form $L_m=p$ \cite{Stanuk1964,Stanuk}.
Using this Lagrangian, the corresponding energy-momentum tensor is (details are given in Appendix in Ref.~\cite{arXiv:1106.1267})
 \begin{equation}
\label{emt_cham_star}
T_i^k=f\left[(\rho+p/c^2)c^2 u_i u^k-\delta_i^k p\right]+\partial_{i}\varphi\partial^{k}\varphi
-\delta_i^k\left[\frac{1}{2}\partial_{\mu}\varphi\partial^{\mu}\varphi-V(\varphi)\right] ~,
\end{equation}
where $\rho$ and $p$ are the mass density and the pressure of the fluid, $u^i$ is the four-velocity.
Henceforth we will use the Lagrangian
\eqref{lagran_cham_star_col} and the  energy-momentum tensor \eqref{emt_cham_star} both when considering static configurations and
in the study of the collapse.
It will be shown  in Appendix \ref{eqs_self_general}
that the process of the collapse permits self-similar motions only in the case of a massless chameleon scalar field,
i.e. when  $V(\varphi)=0$. Bearing this in mind, we restrict ourselves to the consideration  of configurations supported
by the massless chameleon scalar field only.

Now we derive the equations describing static configurations.
In  generalized curvilinear coordinates the static spherically symmetric metric
in the nonrelativistic limit can be written in the form
\cite{Land1}
 \begin{equation}
\label{metric_sphera}
ds^2=e^{\nu(r)}c^2 dt^2-dr^2-r^2d\Omega^2,
\end{equation}
where $d\Omega^2$ is the metric on the unit 2-sphere.
Using this metric, the $r$-component of the covariant conservation law
$T^{\nu}_{\mu; \nu}=0$ (which is the only nonzero component in the static case) is
\begin{equation}
\label{conserv_2_cham_star}
\frac{d p}{d r}=-\frac{1}{2}\rho c^2 \frac{d\nu}{d r}.
\end{equation}
In the nonrelativistic limit,
in turn, the metric function
$\nu$ can be rewritten in terms of the Newtonian gravitational potential
 $\psi$ in the following form \cite{Land1}:
\begin{equation}
\label{metr_newt}
e^{\nu}=1+\frac{2\psi}{c^2}.
\end{equation}
The corresponding equation for
 $\psi$ (the Poisson equation) can be found from the Einstein equations  by using the transition to the nonrelativistic limit
\cite{Land1}. Then, using the
 above expression for $\nu$ and the energy-momentum tensor \eqref{emt_cham_star}, and leaving  terms of order
 $1/c^2$ only, we get
\begin{equation}
\label{poiss_eq}
\frac{1}{r^2}\frac{d}{d r}\left(r^2\frac{d\psi}{d r}\right)=4\pi G f\rho.
\end{equation}
Next, introducing the mass
$$
\frac{d M}{d r}=4\pi f r^2 \rho\,,
$$
one can find from the Poisson equation \eqref{poiss_eq}:
$$
\frac{d \psi}{d r}=\frac{G M(r)}{r^2}.
$$
Using this expression and taking into account
 \eqref{metr_newt}, Eq.~\eqref{conserv_2_cham_star} takes the form
\begin{equation}
\label{conserv_2_cham_star_2}
\frac{d}{d r}\left(\frac{r^2}{\rho}\frac{d p}{d r}\right)=-4\pi G f r^2 \rho.
\end{equation}
This equation differs from the usual equation of hydrostatic equilibrium by  the factor
 $f=f(\varphi)$ reflecting the presence in the system of
 the nonminimal coupling. Correspondingly, a full description of the configuration under consideration requires an additional
  equation for the scalar field $\varphi$. In general form, it can be obtained by varying the Lagrangian
 \eqref{lagran_cham_star_col} with respect to  $\varphi$ as follows:
$$
\frac{1}{\sqrt{-g}}\frac{\partial}{\partial x^i}\left[\sqrt{-g}g^{ik}\frac{\partial \varphi}{\partial x^k}\right]=
L_m \frac{d f}{d \varphi},
$$
where the metric $g_{ik}$ is taken from \eqref{metric_sphera}.
This field equation with the above Lagrangian for the perfect fluid $L_m=p$, and the metric
\eqref{metric_sphera} with \eqref{metr_newt}, gives the following scalar field equation:
\begin{equation}
\label{sf_cham_star}
\varphi^{\prime\prime}+\frac{2}{r}\varphi^\prime=
-p\frac{d f}{d\varphi}~,
\end{equation}
where the prime denotes  differentiation with respect to $r$.  Eqs.~\eqref{conserv_2_cham_star_2} 
and \eqref{sf_cham_star} should be supplemented by an equation of state.
In the case of an isothermal fluid being considered here, we have
\begin{equation}
\label{eos_fluid}
p=K\rho
\end{equation}
with $K=a^2$, where $a$ is the speed of sound in the fluid. Using this equation of state and introducing the dimensionless variables
\begin{equation}
\label{dmls_vars}
e^{-\eta}=\frac{\rho}{\lambda}, \quad
\xi=\frac{r}{L},\quad
\phi=\frac{\sqrt{4\pi G}}{K}\, \varphi,
\end{equation}
where $L=\sqrt{K/(4\pi G \lambda)}$ has dimensions of length,
 $\lambda$ is an arbitrary constant,
Eqs.~\eqref{conserv_2_cham_star_2} and \eqref{sf_cham_star} can be rewritten as follows:
\begin{eqnarray}
\label{conserv_2_cham_star_2_dmls}
&&\frac{1}{\xi^2}\frac{d}{d\xi}\left(\xi^2\frac{d\eta}{d\xi}\right)=f e^{-\eta},
\\
\label{sf_cham_star_dmls}
&&\frac{1}{\xi^2}\frac{d}{d\xi}\left(\xi^2\frac{d\phi}{d\xi}\right)=-e^{-\eta}\frac{d f}{d\phi}.
\end{eqnarray}
In the absence of the scalar field,  i.e. when $f=1$,
only the first equation  remains which
has the following solutions~\cite{Chandra:1939}:
 (i) an
 analytical singular solution in the form $e^{-\eta}=2/ \xi^2$ which diverges at the center of the configuration as
 $\xi\to 0$;
(ii)
numerical regular solutions which satisfy the boundary conditions $\eta=0, \, d\eta/d\xi=0$ at $\xi=0$.
These solutions approach asymptotically, as $\xi \to \infty$,  to the singular solution from (i).
Both sets of solutions imply that  $e^{-\eta}\to 0$ only asymptotically
 as $\xi \to \infty$. That is, such solutions describe only infinite size configurations.

It will  be shown below that the presence of the chameleon scalar field permits the existence not only singular and regular solutions
similar to those described above, but also new
finite size, regular solutions. If we are going to seek such solutions,
it is obvious that a crucial role will be played by the form of the coupling function
 $f$. In the next three subsections we consider several types of solutions. In doing so, we will track  the behavior of the energy density
  $T_0^0$ from \eqref{emt_cham_star}. In the nonrelativistic approximation using the above dimensionless variables
\eqref{dmls_vars},  the total mass density
has the form
\begin{equation}
\label{enrg_dens_stat}
\rho_t \equiv T_0^0/c^2=\lambda\left[f e^{-\eta}+\frac{1}{2}\frac{K}{c^2}\left(\frac{d\phi}{d\xi}\right)^2\right].
\end{equation}
The second term on the right-hand side of this expression is proportional to the square of the ratio of  the speeds of sound and light, $(a/c)^2$.
It is obvious that this term is small compared with the first term everywhere  just except perhaps
the vicinity of the points where both terms tend to zero simultaneously.

Below, we demonstrate a few examples of solutions with different choices
 of the coupling function $f$.  In making the choice of  $f$, two approaches are possible:
  (i)  One assumes $f$ to be an arbitrary function of the scalar field
 $\phi$ whose form is chosen to satisfy the requirement of obtaining the solutions needed.
 As an example, in Sec.~\ref{choice_f_exp} one of the simplest choice of
 $f$ in the form of the exponential function, $f=e^{-\phi}$, is presented. In this case, it is possible to get analytical singular and numerical regular solutions.
 (ii) One can suggest $f$ to be a function of the radial coordinate, $f=f(\xi)$.
 As an example,  we choose below power functions for  $f$ giving either infinite size singular solutions
(see Sec.~\ref{choice_f_pow_sing}) or finite size, regular solutions
(see Sec.~\ref{choice_f_pow_reg}). The solutions for the scalar field  $\phi=\phi(\xi)$ obtained in this case
define the parametric dependence  $f=f(\phi(\xi))$.

 \subsection{Singular and regular solutions with  $f=e^{-\phi}$}
\label{choice_f_exp}

In this section, we seek a solution of the system
  \eqref{conserv_2_cham_star_2_dmls}-\eqref{sf_cham_star_dmls} in the particular case when
 the coupling function $f$ is chosen in the form
\begin{equation}
\label{choice_f_exp_sing}
f=f_0 e^{-\phi}.
\end{equation}
The parameter $f_0>0$ can be absorbed by introducing the rescaling $\sqrt{f_0} \xi \to \xi$
that allows to put
$f_0=1$ in further calculations. But we will bear in mind  that
 $f_0$ can be always restored in final expressions by using the above rescaling.
The above choice of $f$ allows to find the following analytical solutions
\begin{equation}
\label{sol_stat_f_exp}
e^{-\eta}=\frac{A \exp{(-B/\xi)}}{\xi}, \quad
\phi=\ln{(A\, \xi)}-\frac{B}{\xi},
\end{equation}
where $A, B$ are integration constants. These solutions describe a configuration which
 is singular at the origin of coordinates, where, as
 $\xi\to 0$, the mass density
from \eqref{enrg_dens_stat} diverges as $1/\xi^2$,
and regular asymptotically,
 as $\xi\to \infty$, where $\rho_t \to 0$.

The coupling function $f$ from \eqref{choice_f_exp_sing} permits also the existence of regular solutions which can be found
numerically.
In this case we can choose the parameter $\lambda$ from \eqref{dmls_vars} to be the central density of the fluid, $\lambda=\rho_c$.
With this normalization the system \eqref{conserv_2_cham_star_2_dmls}-\eqref{sf_cham_star_dmls} takes the form
\begin{eqnarray}
\label{eq_eta_f_exp}
&&\frac{1}{\xi^2}\frac{d}{d\xi}\left(\xi^2\frac{d\eta}{d\xi}\right)=e^{-\eta-\phi},
\\
\label{eq_phi_f_exp}
&&\frac{1}{\xi^2}\frac{d}{d\xi}\left(\xi^2\frac{d\phi}{d\xi}\right)=e^{-\eta-\phi}.
\end{eqnarray}
We will look for  a solution of this system
which satisfy the following boundary conditions at $\xi=0$:
\begin{equation}
\label{bound_cond_reg_f_exp}
\eta=0, \quad \frac{d\eta}{d\xi}=0, \quad \phi =\phi_0=\text{const}, \quad \frac{d\phi}{d\xi}=0.
\end{equation}
Taking these conditions into account, one can show that solutions of Eqs.~\eqref{eq_eta_f_exp} and \eqref{eq_phi_f_exp} 
are related through
 $\phi=\eta+\phi_0$. This allows the possibility of reducing the system
 \eqref{eq_eta_f_exp}-\eqref{eq_phi_f_exp}  to one equation:
$$
\frac{1}{\xi^2}\frac{d}{d\xi}\left(\xi^2\frac{d\eta}{d\xi}\right)=e^{-2\eta-\phi_0},
$$
which, after the rescalings
 $\xi = \zeta\, e^{\phi_0/2}/\sqrt{2}, \,\,\eta = \theta/2$,
takes the form
\begin{equation}
\label{eq_eta_f_exp_alone}
\frac{1}{\zeta^2}\frac{d}{d\zeta}\left(\zeta^2\frac{d\theta}{d\zeta}\right)=e^{-\theta}.
\end{equation}
This equation is an analogue of the Lane-Emden equation for the isothermal fluid without a scalar field
\cite{Chandra:1939}.
Thus, Eq.~\eqref{eq_eta_f_exp_alone} [or equivalently Eqs.~\eqref{eq_eta_f_exp} and \eqref{eq_phi_f_exp}] 
together with the boundary conditions
\eqref{bound_cond_reg_f_exp} describes the regular
 isothermal gas sphere in the presence of the chameleon scalar field. Without the scalar field,
 such configuration has the central density $\rho_c$. In the presence of the scalar field,
 it follows from Eq.~\eqref{enrg_dens_stat} 
 that the central density of the configuration under consideration is now defined by the value of the
function $f$ at $\xi=0$ as well. In the case when the coupling function has the form  \eqref{choice_f_exp_sing},
we have from  \eqref{enrg_dens_stat}:
$\rho_{tc}=\rho_c e^{-\phi_0}$. It implies that the configuration under consideration will have a greater or smaller concentration of matter
at the center depending on the sign of $\phi_0$,
and correspondingly different distributions of the mass density along the radius.
The size of such configuration, as in the case without the  scalar field, will also be infinite.
The corresponding asymptotic solutions of the system  \eqref{eq_eta_f_exp}-\eqref{eq_phi_f_exp} are:
$$
\eta=\phi \sim \ln{\xi} \quad \Rightarrow \quad \rho_t \sim \xi^{-2} \quad \text{as} \quad\xi\to \infty.
$$

As in the case of an isothermal fluid without a scalar field, here we will consider a configuration
 embedded in an external medium of
pressure $p_{e}$. For such static configurations without a scalar field, in the works
of Ebert \cite{Ebert:1955} and Bonnor \cite{Bonnor:1956}, the issue of their stability was investigated.
Following these works and taking into account the equation of state from \eqref{eos_fluid},
we represent the central density of the fluid
 $\rho_c$ of the configuration under consideration
as follows (see also Ref.~\cite{Hunter:1977}):
\begin{equation}
\label{rho_c_fluid}
\rho_c=\frac{p \,e^{\eta(\xi)}}{a^2}.
\end{equation}
Then, taking into account the total pressure $p_t$, which includes both the pressure of the fluid and
the contribution from
the scalar field, and follows from
 \eqref{emt_cham_star},
$p_{t}=f p+ (d\varphi/d r)^2/2$,  in terms of the  dimensionless variables
 \eqref{dmls_vars}, we have
$$
p=\frac{1}{f}\left[p_{t}-\frac{a^2 \rho_c}{2}\left(\frac{d\phi}{d\xi}\right)^2\right].
$$
Substituting this expression in
 \eqref{rho_c_fluid} and taking into account that
 at the outer boundary of the configuration, where
$\xi=\xi_{e}$, the total pressure is equal to the external pressure,
i.e.
$p_t(\xi_e)=p_e$, we finally have
\begin{equation}
\label{rho_c_fluid_2}
\rho_c=\frac{p_{e}}{a^2}\left[f[\phi(\xi_{e})]\, e^{-\eta(\xi_e)}+\frac{1}{2}\left(\frac{d\phi}{d\xi}\right)^2_{\xi=\xi_e}\right]^{-1}.
\end{equation}
 Then one can find that the total mass of the configuration is
$$
M=\frac{a^4}{G(4\pi G p_{e})^{1/2}}\,\xi_{e}^2 \left[f[\phi(\xi_{e})]\,e^{-\eta(\xi_{e})}
+\frac{1}{2}\left(\frac{d\phi}{d\xi}\right)^2_{\xi=\xi_e}
\right]^{1/2}\,
\left(\frac{d\eta}{d\xi}\right)_{\xi=\xi_{e}}.
$$
The maximum possible $M$, given $p_e$, or the maximum possible $p_e$, given $M$, follows from the fact that the
function $\xi^2 \sqrt{f[\phi(\xi)]\,e^{-\eta(\xi)}+1/2 \phi^{\prime 2}}\, d\eta/d\xi$ attains a maximum value at some $\xi=\xi_{max}$.
In the case when
 $f$ is chosen in the form
 \eqref{choice_f_exp_sing}, the maximum value of $\xi$ is $\xi_{max}=5.235 \,e^{\phi_0/2}$.
One can see that this value depends on the initial value of the field
 $\phi_0$, and differs from the case without a scalar field when
 $\xi_{max}=6.451$ \cite{Bonnor:1956, Hunter:1977}. The maximum value
 of the radial coordinate, $\xi_{max}=5.235\, e^{\phi_0/2}$, corresponds to the mass of the cloud
having the critical value
\begin{equation}
\label{M_crit_f_exp}
M_{\text{crit}}=0.52\frac{a^4}{G^{3/2}}p_{e}^{-1/2}.
\end{equation}
This value is factor of $0.44$ smaller than the critical Bonnor-Ebert sphere. Notice that the critical mass does not depend on the
initial value of the scalar field
 $\phi_0$ and is determined  just by the external pressure
 $p_e$, as in the case of
 the critical Bonnor-Ebert sphere. The relation between the central density and the total density at the outer boundary of the
 configuration,  $\rho_c/\rho_t=f^{-1}\,e^{\eta}$, corresponding to this critical mass, can be written as
 $\rho_c/\rho_t(\xi_{max})\simeq 19.724 \,e^{\phi_0}$.
Then for clouds with masses less than $M_{\text{crit}}$ two types of configurations are possible: a larger stable one
with a range of density less than $19.724 \,e^{\phi_0}$, and a  smaller unstable one with a range of density greater than $19.724 \,e^{\phi_0}$.
For both cases $\xi_e<\xi_{max}$. Configurations with $\xi_e>\xi_{max}$ are always unstable, and with  $\xi_e=\xi_{max}$ are
marginally stable.

\begin{figure}[t]
\centering
  \includegraphics[height=9cm]{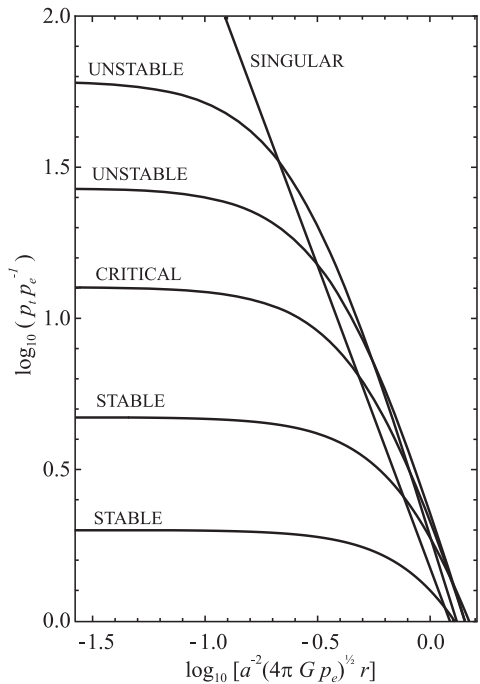}
\caption{Total pressure distributions of bounded isothermal spheres in the presence of the chameleon scalar field
with the coupling function $f$ taken from \eqref{choice_f_exp_sing}.
All nonsingular curves are plotted at the initial value of $\phi_0=0$. The curve marked ``critical'' corresponds
to the configuration with the critical mass given by Eq.~\eqref{M_crit_f_exp}. The spheres which are less centrally
concentrated that the critical configuration are gravitationally stable; those which are more centrally
concentrated are gravitationally unstable. In the limit of infinite central concentration, the latter spheres approach the
singular solution.
}
\label{fig_dens_stat}
\end{figure}

In the limit of configurations with infinite central density
described by the singular solutions \eqref{sol_stat_f_exp},
the unstable equilibria approach the singular sphere which has the following density and mass distributions:
\begin{equation}
\label{sing_sol_dim}
\rho_t(r)=\frac{a^2 f_0}{4\pi G} r^{-2}, \quad M(r)=\frac{a^2}{f_0^{1/2} G}\,r.
\end{equation}
This singular solution, truncated at a boundary pressure $p_e$, has a total radius $R$ and a total mass $M(R)$ given by
\begin{equation}
\label{sing_M_R}
R=\frac{a^2 f_0^{1/2}}{(4\pi G)^{1/2}}p_e^{-1/2}, \quad
M(R)=\left(\frac{1}{4\pi}\right)^{1/2}\frac{a^4}{G^{3/2}}\,p_e^{-1/2},
\end{equation}
which are, respectively, factors of $0.58 f_0^{1/2}$ and $0.24$ differ from the critical Bonnor-Ebert sphere.
Here we restored the parameter
 $f_0$ from \eqref{choice_f_exp_sing} whose value allows varying the radius of the configuration for the given mass.
The plots of  total pressure
distributions of several configurations described above are shown in Fig.~\ref{fig_dens_stat}.

 \subsection{Singular solutions with    $f=f(\xi)$}
\label{choice_f_pow_sing}

In this and the next subsections, we seek solutions of the system
  \eqref{conserv_2_cham_star_2_dmls}-\eqref{sf_cham_star_dmls} starting not from a straightforward choice of
  the specific form
 of the coupling function $f$, but by choosing  an appropriate  form of the product
 $f e^{-\eta}$, defining the total density \eqref{enrg_dens_stat}.
In  paper \cite{arXiv:1106.1267}, we  already found
the analytical solutions for a massless scalar field
in a particular  nonrelativistic case of an incompressible fluid.
Below we will obtain analytical solutions for the case of the isothermal fluid being considered here.

 Choose first the above product in the form
\begin{equation}
\label{choice_fe_sing}
  f e^{-\eta}=\frac{B}{\xi^p},
\end{equation}
where $B, p$ are arbitrary positive constants. This choice gives the positive-definite total mass density
 $\rho_t$ from \eqref{enrg_dens_stat} which is singular at the origin of coordinates.
 The choice \eqref{choice_fe_sing} allows to find the general solution of Eq.~\eqref{conserv_2_cham_star_2_dmls} 
 in the form
$$
\eta=C_1+\frac{C_2}{\xi}+\frac{B \xi^{2-p}}{(p-2)(p-3)}, \quad
p\neq 2,3.
$$
In the case when
$p=2,3$, we have
$$
\eta_2=C_1+\frac{C_2}{\xi}+B \ln{\xi}, \quad
\eta_3=C_1+\frac{C_2}{\xi}-\frac{B}{\xi} \left(1+\ln{\xi}\right).
$$
Here $C_1, C_2$ are integration constants.
The scalar field equation \eqref{sf_cham_star_dmls}, in turn, can also be  integrated analytically giving
the expression for $\phi^{\prime 2}$.

Of special interest is the case of  $p=2$. In this case one can get a solution similar to the singular solution without a
scalar field. Indeed, choosing the
integration constants as $C_2=0$  and $C_1=\ln {(1/2)}$, we obtain the following solutions:
\begin{equation}
\label{fe_sing_sol_2}
\eta_2=\ln{\frac{\xi^B}{2}}, \quad \phi^{\prime 2}_2=\frac{B(2-B)}{\xi^2}, \quad f_2=\frac{B}{2}\xi^{B-2}.
\end{equation}
In the case when $B=2$ we return to the singular solution without a scalar field from (i) [see after Eq.~\eqref{sf_cham_star_dmls}].
The values of the constant  $B\neq 2$ imply that the scalar field changes along the radius
that provides  changes in the distribution of the mass density from \eqref{enrg_dens_stat}.

It can be also shown that, putting $B=1, p=2$ in \eqref{choice_fe_sing}, solutions \eqref{fe_sing_sol_2} correspond
to the singular solutions from Sec.~\ref{choice_f_exp}
if one chooses the parameters from expressions \eqref{sol_stat_f_exp} to be $A=2, B=0$.
Thus we have
$$
f=\frac{1}{2\xi}, \quad \phi=\ln{(2\xi)}, \quad  e^{-\eta}=\frac{2}{\xi}.
$$
It is obvious from two first expressions that the parametric dependence
 $f=f(\phi(\xi))$ gives the expression
 $f=e^{-\phi}$ coinciding with the coupling function from
 \eqref{choice_f_exp_sing}. In Sec.~\ref{collaps_sing} below,  
 a self-similar motion of matter with such choice of $f$ will be considered.

 \subsection{Regular solutions with  $f=f(\xi)$}
\label{choice_f_pow_reg}

Bearing in mind that we  seek finite size, regular solutions, let us try to choose such an expression for
 $f e^{-\eta}$ that provides the required solutions.
In doing so, we will look for such solutions of Eqs.~\eqref{conserv_2_cham_star_2_dmls} and \eqref{sf_cham_star_dmls}
 which have a value of the total mass density
 $\rho_t$ from \eqref{enrg_dens_stat} at the center of the configuration, at
 $\xi= 0$, equal to the central density of the fluid,
 $\lambda=\rho_c$, i.e. $\rho_{t0} =\rho_c$, and at the outer boundary of a cloud,
at $\xi=\xi_1$, we require that
 $\rho_{t1} = 0$. One of the simplest variants is to take the following power-law dependence
\begin{equation}
\label{choice_fe}
  f e^{-\eta}=1+\alpha \xi^\beta,
\end{equation}
where $\alpha, \beta$ are arbitrary constants. Taking all the above into account, we  choose
$\alpha$ to be negative, and  $\beta>0$. Substituting  expression
 \eqref{choice_fe} in Eqs.~\eqref{conserv_2_cham_star_2_dmls} and \eqref{sf_cham_star_dmls}, 
 we find their solutions in the form
\begin{eqnarray}
\label{eta_z}
\eta&=&\frac{1}{6}\xi^{2}+\frac{\alpha}{(\beta+2)(\beta+3)}\xi^{2+\beta},
\\
\label{phi_pr_z}
\left(\frac{d\phi}{d \xi}\right)^{ 2}&=&-\frac{\xi^2}{9}
\left\{1+\frac{6\alpha}{\beta+3}\xi^\beta\left[
1+\frac{3\beta(\beta+3)}{\beta+4}\frac{1}{\xi^2}+\frac{3\alpha}{2(\beta+3)}\xi^\beta
\right]
\right\}.
\end{eqnarray}

Next, from expression \eqref{choice_fe}, let us determine
the point $\xi=\xi_1$ in which
 $ f e^{-\eta}=0$: $\xi_1=(-\alpha)^{-1/\beta}$. Requiring that in this point the derivative
 $\phi^{\prime}$ also be equal to zero, we substitute this value of
 $\xi_1$ in Eq.~\eqref{phi_pr_z} and find the corresponding value of  $\alpha$:
$$
\alpha=-\left\{\frac{\beta(\beta+4)}{18(\beta+3)^2}
\right\}^{\beta/2}.
$$
Using this expression, we get
$$
\xi_1=\left\{\frac{\beta(\beta+4)}{18(\beta+3)^2}
\right\}^{-1/2}.
$$
This point can be interpreted  as the outer boundary of the configuration, where the total density from 
Eq.~\eqref{enrg_dens_stat} is equal to zero.
 In this case the spheres described by Eqs.~\eqref{conserv_2_cham_star_2_dmls} and \eqref{sf_cham_star_dmls}  
 are a one-parameter family,
with the size of $\xi_1$ depending only on one parameter $\beta$. As $\beta \to \infty$, the value
$\xi_1 \to 3\sqrt{2}$, and as $\beta \to 0$, we have $\xi_1 \sim 1/\sqrt{\beta} \to \infty$.
Thus, by changing the value of the parameter $\beta$, we can change the size of the configuration
in a wide range
at the given $\rho_c$ and $K$.

Note that the scalar field energy density which is proportional to
 $\phi^{\prime 2}$ from
  \eqref{phi_pr_z}, can take both positive and negative values along the radius depending on the value of the
  parameter  $\beta$. This is related to the structure of the scalar field equation
\eqref{sf_cham_star} which, at positive pressure  $p$ and a specific choice of  $f$,
can  effectively correspond to a scalar field equation
for a ghost scalar field. Since the form of $f=f(\phi(\xi))$ is  defined parametrically through
the solution for the scalar field
 $\phi=\phi(\xi)$, and depends on the value of the parameter $\beta$, then a situation may occur where
 the scalar field equation \eqref{sf_cham_star} demonstrates the ghostlike behavior with $\phi^{\prime 2}<0$.
However, since in the nonrelativistic limit the main contributions to the energy density
 \eqref{enrg_dens_stat} come from the first term on the right-hand side, which is always positive, then
 the total energy density remains positive as well.

Returning to the dimensional variables from \eqref{dmls_vars}, the radius of the configuration under consideration is
given by
$$
R=\left(\frac{K}{4\pi G \rho_c}\right)^{1/2}\xi_1.
$$
In Sec.~\ref{collaps_sing}, where we will consider the process of the collapse of the singular sphere from 
Sec.~\ref{choice_f_exp}, the speed of sound is chosen to be
 $a=0.2\, \text{km s}^{-1} $. If we use the same value in the estimation of the size of the regular configuration which is
  considered here, we have
$$R=2.18\times 10^7 \frac{\xi_1}{\sqrt{\rho_c}} \,\,\text{cm},$$
where $\rho_c$ is the central density of the configuration in $\text{g cm}^{-3}$.

It is seen from the obtained results that the isothermal fluid in the presence of the nonminimally coupled scalar field with
the coupling function $f$  defined through Eq.~\eqref{choice_fe} permits the existence of finite size, regular solutions.
This situation differs from the classical problem without a scalar field when
infinite size solutions are the only possible ones \cite{Chandra:1939}.

Let us now consider the issue of the stability of the solutions obtained.
Following \cite{Zeld}, using Eq.~\eqref{conserv_2_cham_star_2_dmls}, we introduce a quantity
$$
\mu_1\equiv \xi_1^2 \left(\frac{d\eta}{d\xi}\right)_{\xi=\xi_1}=\int_0^{\xi_1} f \xi^2 e^{-\eta}d\xi.
$$
Because of the nonrelativistic character of the problem, we neglect the influence of the scalar field.
Next, using expression \eqref{enrg_dens_stat}, we define the average density as
$$
\bar{\rho}_t=\frac{M}{(4/3)\pi R^3}=\frac{M}{(4/3)\pi L^3 \xi_1^3}.
$$
For the density concentration, we have
$$
\frac{\rho_c}{\bar{\rho}_t}=\frac{\xi_1^3}{3\mu_1}.
$$
By combining two last expressions, we  find
$$
\rho_c=\frac{M}{4\pi L^3 \mu_1}.
$$
Taking into account the expression for
 $L$ from \eqref{dmls_vars}, we obtain
\begin{equation}
\label{rho_c_stab}
\rho_c=\frac{K^3}{4\pi G^3}\left(\frac{\mu_1}{M}\right)^2.
\end{equation}
One can see  from this equation that  $\rho_c$ drops with increasing $M$, which
is the abnormal behavior for a stable star in equilibrium.
Thus, regardless of the value of the parameter
 $\beta$, the choice of the coupling function $f$ in the form
\eqref{choice_fe} gives only unstable configurations.

\section{Self-similar collapse of the singular sphere}
\label{collaps_sing}

In this section we consider the spherical gravitational collapse of the singular configuration investigated in 
Sec.~\ref{choice_f_exp}.
 Such problems have been considered frequently
  for polytropic spheres both
 in relativistic and nonrelativistic cases. In doing so, two main approaches are being used:
(i) One solves the system of hydrodynamic partial differential equations;
(ii) A self-similar motion of  polytropic matter is considered.

Here we study self-similar motions of the matter of the singular configuration considered in Sec.~\ref{choice_f_exp}. 
In the absence of the
 chameleon scalar field, such problem for an isothermal sphere have been considered by many authors
 (see the Introduction). We obtain below
 the similarity solutions  describing the collapse of the singular sphere
 in the presence of the chameleon scalar field. These solutions will be compared to the known solutions found in Shu's paper
 \cite{Shu:1977uc}.

In Appendix \ref{eqs_self_general} we derived the following set of nonrelativistic
 hydrodynamic equations which take into account  the influence of
 a chameleon scalar field:
\begin{eqnarray}
\label{eq_cont_2_m}
\frac{\partial M}{\partial t}+4\pi f r^2 \rho v&=&0,\\
\label{mass_fun_2_m}
\frac{\partial M}{\partial r}&=&4\pi f r^2 \rho,\\
\label{eq_eul_3_m}
\frac{\partial v}{\partial t}+v \frac{\partial v}{\partial r}&=&-\frac{1}{\rho}\frac{\partial p}{\partial r}-
\frac{G M}{r^2},\\
\label{sf_cham_star_col_1_m}
\frac{\partial^2 \varphi}{\partial r^2}+\frac{2}{r}\frac{\partial \varphi}{\partial r}-\frac{\partial V}{\partial \varphi}&=&
-p\frac{\partial f}{\partial\varphi}.
\end{eqnarray}
Here $M=M(r,t)$ is the total mass inside radius $r$ at time $t$, $v$ the radial velocity, $p$ the pressure,
and $\rho$ the density.
If the coupling function is chosen in the form  $f=e^{-\phi}$, then
the corresponding system of  self-similar Eqs.~\eqref{eq_fin_1}-\eqref{eq_fin_3} from Appendix \ref{eqs_self_general}
can be written as follows:
\begin{eqnarray}
\label{eq_fin_1_exp}
&&\frac{d\alpha}{d x}=\frac{\alpha}{(x-u)^2-1}\left[e^{-\phi}\alpha-2\frac{x-u}{x}+(x-u)\frac{d\phi}{dx}\right](x-u),\\
\label{eq_fin_2_exp}
&&\frac{d u}{d x}=\frac{1}{(x-u)^2-1}\left[e^{-\phi}\alpha(x-u)-\frac{2}{x}+\frac{d\phi}{dx}\right](x-u),\\
\label{eq_fin_3_exp}
&&\frac{d^2\phi}{d x^2}+\frac{2}{x}\frac{d\phi}{d x}=
\alpha e^{-\phi}.
\end{eqnarray}
The similarity variables involved were obtained from expressions
 \eqref{self_variab} in the case when $\gamma=n=1$, $k=K=a^2$.
This gives
\begin{align}
\label{self_variab_f_exp}
\begin{split}
x&=\frac{r}{a\, t},\quad v(r,t)=a u(x), \quad \rho(r,t)=\frac{\alpha(x)}{4\pi G t^2}, \\
p(r,t)&=\frac{a^2}{4\pi G t^{2}}\,\alpha(x), \quad M(r,t)=\frac{a^{3} t}{G}\,m(x), \quad
\varphi(r,t)= \frac{a^2}{\sqrt{4\pi G}}\,\phi(x),
\end{split}
\end{align}
where $t=0$ defines the instant when the mass of the core, $M(0,t)$, is zero. The instant $t=0$ corresponds to
the instant of core formation for the collapse problem. In the case of collapse the variables $t, x, m$ are positive
while $u$ is negative.

In terms of the above similarity variables we have the following expression for the mass function $m(x)$ from \eqref{mass_fun_algeb}:
\begin{equation}
\label{mass_fun_algeb_f_exp}
m(x)=e^{-\phi}x^2 \alpha (x-u).
\end{equation}
Thus we have the system of three Eqs.~\eqref{eq_fin_1_exp}-\eqref{eq_fin_3_exp} and relation \eqref{mass_fun_algeb_f_exp}
describing the self-similar collapse of the configuration under consideration.

An exact analytic solution of Eqs.~\eqref{eq_fin_1_exp}-\eqref{eq_fin_3_exp} is given by the static solutions
\eqref{sol_stat_f_exp}
 in the particular case of $B=0$:
\begin{equation}
\label{stat_sols_ini_f_exp}
u=0, \quad \alpha=\frac{A}{x}, \quad \phi=\ln{(A x)},
\quad m=x.
\end{equation}
In dimensional units, these solutions correspond to the time-independent singular solution
\eqref{sing_sol_dim} with $f_0=1$. As in the case of the problem without a scalar field considered
 by Shu in \cite{Shu:1977uc}, in our case
this singular solution is the only hydrostatic solution allowing self-similarity. Following \cite{Shu:1977uc},
we will use this solution as the ``initial state'' when the process of collapse starts.

Below, we will seek numerical solutions of the system
 \eqref{eq_fin_1_exp}-\eqref{eq_fin_3_exp}. However, this numerical study
needs some caution because the solution can pass through critical points, as well as in the case of the absence of a scalar field \cite{Shu:1977uc}.
The existence of the critical points assumes that the denominator of Eqs.~\eqref{eq_fin_1_exp} and \eqref{eq_fin_2_exp} vanishes:
$$
(x-u)^2-1=0.
$$
Substituting this expression in Eqs.~\eqref{eq_fin_1_exp}-\eqref{eq_fin_3_exp}, one can obtain the following set of equations:
\begin{align}
\label{eqs_crit_point}
\begin{split}
&\frac{d\phi}{d x}+e^{-\phi}\alpha -\frac{2}{x}=0, \\
&\frac{d^2\phi}{d x^2}+\frac{2}{x}\frac{d\phi}{d x}=
\alpha e^{-\phi}.
\end{split}
\end{align}
Solutions of these equations make
 the numerators of  Eqs.~\eqref{eq_fin_1_exp} and \eqref{eq_fin_2_exp} equal to zero, thereby
 providing the regularity of the solutions  at the critical  points.
The particular solutions of the system \eqref{eqs_crit_point} are
$$
\alpha=2 e^{2/x}, \quad \phi=\frac{2}{x}+2\ln{x}.
$$
One can show that these solutions do not satisfy
 the original Eqs.~\eqref{eq_cont_2_m}-\eqref{sf_cham_star_col_1_m}.
Henceforth we will only consider  the solutions which do not pass through the
 critical points.

\begin{figure}[t]
\centering
  \includegraphics[height=9cm]{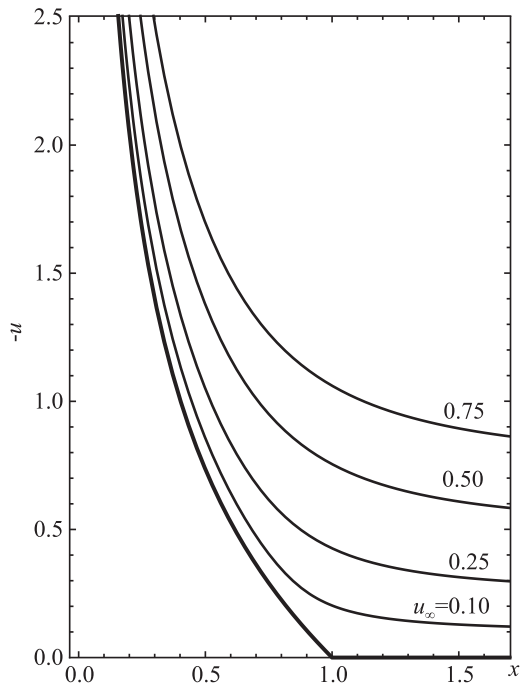}
\caption{Similarity solutions for the velocity $u$ with different initial values of $u_{\infty}$.
The thick solid curve gives the collapse solution for the singular sphere \eqref{bound_distr_fluid}
which is initially hydrostatic (``expansion-wave collapse solution'').
}
\label{fig_vels}
\end{figure}

To find a solution of Eqs.~\eqref{eq_fin_1_exp}-\eqref{eq_fin_3_exp}, it is necessary to
specify the corresponding boundary conditions. When considering a problem of collapse,
 it is reasonable to assume that the fluid velocity is negligible
at the ``initial instant,'' i.e. that $u\to 0$ as $x\to \infty$.
Then one can show that
 solutions of the system \eqref{eq_fin_1_exp}-\eqref{eq_fin_3_exp} which have this property have
the following asymptotic behavior:
\begin{equation}
\label{bound_col}
\alpha \sim A/x, \quad u \sim u_{\infty} e^{1/(2 x^2)}, \quad \phi \sim \ln{(A x)},
\quad m \sim x \quad \text{as} \quad x \to \infty,
\end{equation}
where $u_{\infty}<0$ is a value of the velocity $u$
as $x\to \infty$.
The collapse begins
 from the initial inhomogeneous distribution of the matter in the form
\begin{equation}
\label{bound_distr_fluid}
\rho_t(r,0)=\frac{a^2}{4\pi G} r^{-2},
\end{equation}
which corresponds to the singular static sphere from \eqref{sing_sol_dim}.
This sphere being initially in  unstable hydrostatic equilibrium,
can spontaneously generate inflow
at every radii when $u_{\infty}\neq 0$.
Several solutions of the system \eqref{eq_fin_1_exp}-\eqref{eq_fin_3_exp} with different initial values of the velocity $u_{\infty}$
are presented in Fig.~\ref{fig_vels}.

Shu starts his similarity solutions with the initial singular state of the type
 \eqref{bound_distr_fluid} using as initial conditions the asymptotic expressions for the velocity
 $u \sim -(D-2)/x$ and for the initial density distribution
$\rho(r,0)\sim D r^{-2}$ depending on the value of the arbitrary parameter
 $D$. In the limiting case $D\to 2$, Shu returns to the singular sphere which is the solution of
 static equations.
As a result, he obtains
the limiting self-similar solutions,  called the
``expansion-wave collapse solutions.''
In our case we also have similar solution shown in Fig.~\ref{fig_vels} by the thick solid line.
We will describe this solution in detail in Sec.~\ref{exp_wave_sols}.

Let us now consider the behavior of collapsing solutions of the system  \eqref{eq_fin_1_exp}-\eqref{eq_fin_3_exp}
in the limit of  $x \ll 1$. Assuming here supersonic flow, $u^2 \gg 1$,
we obtain the following solutions near the origin:
\begin{equation}
\label{sols_origin_f_exp}
m \to m_0, \quad \alpha \to \frac{q}{x^{3/2}}\,e^{-b/x}, \quad
u\to -\frac{q+\sqrt{q^2+2 b}}{x^{1/2}}, \quad \phi\to-\frac{b}{x}.
\end{equation}
Here $b, q$ are constants whose value is determined  from the solution, and depends on
 $u_{\infty}$, and $m_0=q\left(q+\sqrt{q^2+2 b}\right)$
is the value of the mass at the center. When $b=0$, i.e. in the absence of the scalar field, we return to the known solutions
from Ref.~\cite{Shu:1977uc}:
$
m \to m_0, \, \alpha \to (m_0/2 x^3)^{1/2}, \,
u\to -(2 m_0/x)^{1/2}.
$

Following \cite{Shu:1977uc}, let us calculate
 the ratio of the gravitational acceleration to the pressure gradient acceleration for the fluid:
$$
\frac{G M}{r^2}:\frac{a^2}{\rho}\frac{\partial \rho}{\partial r}=
\frac{m}{x^2}:\frac{d\ln{\alpha}}{d x} \to \frac{m_0}{b} \quad \text{as} \quad x\to 0,
$$
which tends to the constant
 at finite $t$ as $r\to 0$. This situation differs from
Shu's result, where the above relation diverges that is quite expected behavior for the collapse problem.
The obtained result is obviously related to the presence in the system of the scalar field
whose behavior becomes dominant in the limit $x\to 0$. In this case one can suppose that
it is needed to perform a more correct study of the problem assuming the presence of the scalar field
in the expression for the energy density
 $T_0^0$ from \eqref{comp_emt_01} which is not  taken into account in the present model
because of its nonrelativistic character.

The constant $m_0$, which represents the reduced mass of the core, is expressed through the above constants $b, q$
which, in turn, are defined implicitly by the constant $u_{\infty}$ from \eqref{bound_col}, but not by $A$ which,
as it is obvious from the structure of Eqs.~\eqref{eq_fin_1_exp}-\eqref{eq_fin_3_exp}, does not influence the solutions.
The values of $b, q$, and correspondingly $m_0$ can be found  by numerical integration with
the starting conditions \eqref{bound_col} determined at large $x$.
Numerical calculations show the following linear dependence which is approximately valid at
 $|u_{\infty}|<10$:
$m_0 \simeq 0.491+|u_{\infty}|$.

\subsection{Expansion-wave collapse solution}
\label{exp_wave_sols}

Following Shu \cite{Shu:1977uc}, in this section we consider similarity solutions starting from the initial static
singular distribution  of matter in the form   \eqref{bound_distr_fluid}. Then, letting the initial velocity
$u_{\infty}$ from \eqref{bound_col} equal to zero, the solutions  obtained below can be interpreted as follows:
we start from the initially unstable singular sphere  \eqref{bound_distr_fluid} with zero initial velocity
 $u$ at the instant $t=0$. The collapse begins at the central region as a small perturbation of density
 spreading in the external envelope at the speed of sound $a$ up to the outer boundary of the configuration,
 $r=R$, defined by Eq.~\eqref{sing_M_R}. Behind the front of this
 expansion wave, the motion of the matter is described by Eqs.~\eqref{eq_fin_1_exp}-\eqref{eq_fin_3_exp} 
 whose solutions are  sought in the range
 $0^+ \leq x \leq 1$. The value $x=1$ corresponds to the outer boundary of the configuration where
 $r=R$. The time required for initiation of the collapse  over the whole configuration is
 $T=R/a$. During the collapse,  in the neighborhood of the center of the configuration, the accumulation of mass   from
 the initial value  $m_0 \simeq 0.491$ to  $m=1$ takes place.
 The latter value represents
 a total reduced mass contained inside  $x=1$ and equal to the original equilibrium value
from \eqref{stat_sols_ini_f_exp}. The physical mass $M$ contained in the expansion wave and in the core increases linearly
with time according to  Eq.~\eqref{self_variab_f_exp}. As well as in Shu's work, the mass contained in the core at any instant
$t$ always comprises about $49\%$ of the total mass contained within $r=a t$.

Let us now illustrate the process of collapse described above by using  numerical calculations with a specific choice of the value of
the speed of sound, $a=0.2\, \text{km s}^{-1}$ \cite{Shu:1977uc}. In Ref.~\cite{Shu:1977uc} the external pressure was chosen to be
$p_e/k=1.1 \times 10^5\, \text{cm}^{-3}\,\text{K}$
(here $k$ is the Boltzmann constant) which defines the size and the total mass of the configuration as
 $R=1.6\times 10^{17}\, \text{cm}\simeq 0.052\, \text{pc}$ and $M=0.96 M_{\odot}$, respectively.
 It allowed Shu to identify such an object as a Bok globule embedded in an H~II region,
 and to use it as an initial state from which the  process of collapse started.
 If we want to get the same value of the mass as in \cite{Shu:1977uc},
 then by comparing our expressions for the size of the configuration and its mass,
 given by Eq.~\eqref{sing_M_R},  with the corresponding expressions from \cite{Shu:1977uc}, given by Eq.~(3), it can be seen that
 in our case one needs to take the pressure $p_e$   8 times smaller than Shu's value. It will lead, in turn, to the growth
 of radius $R$ which is now $\sqrt{8}$ times larger than the value from \cite{Shu:1977uc}.

\begin{figure}[t]
\begin{minipage}[t]{.49\linewidth}
  \begin{center}
  \includegraphics[width=6.4cm]{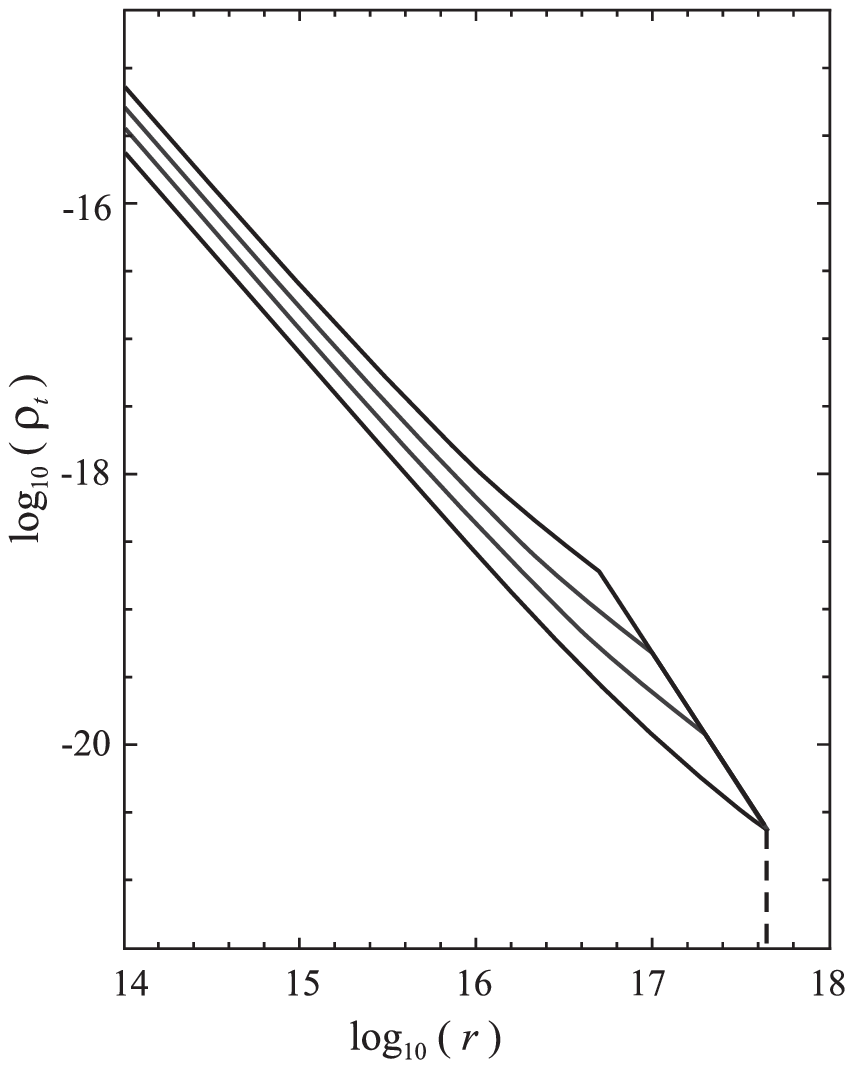}
  \end{center}
\end{minipage}\hfill
\begin{minipage}[t]{.49\linewidth}
  \begin{center}
  \includegraphics[width=6cm]{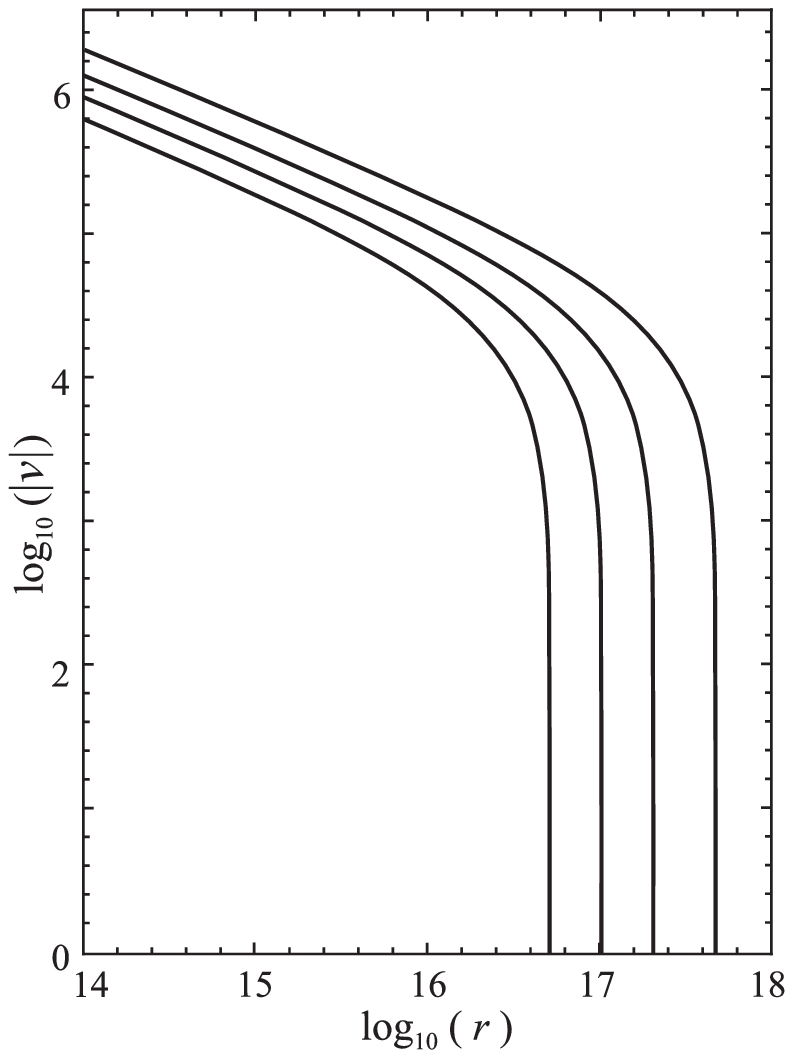}
  \end{center}
\end{minipage}\hfill
  \caption{Expansion-wave collapse solution for a  $0.96 M_{\odot}$ singular sphere with $a=0.2\, \text{km s}^{-1}$
  and $p_e/k=1.4 \times 10^4\, \text{cm}^{-3}\,\text{K}$. The initial radius of the outer boundary is shown
  on the left panel  by the
  vertical dashed line. The total density $\rho_t$ and the velocity $-v$  profiles at
  $t=0.25, 0.50, 1.00,$ and $2.30 \times 10^{13}\, \text{s}$  are shown
   (from top to bottom on the left panel, and from left to right on the right panel, respectively).
   The dimensions of $r, \rho_t$, and $v$ are
  cm, $\text{g}\, \text{cm}^{-3}$, and cm $\text{s}^{-1}$.}
    \label{fig_dens_vels_col}
\end{figure}

The collapse begins  in the neighborhood of the center of the configuration
 at the instant $t=0$. The expansion wave moving through the configuration with the speed of sound
 $a$, reaches the outer boundary
 at  $R=4.5\times 10^{17}\, \text{cm}\simeq 0.147\, \text{pc}$
 (that is about 1000 times the distance of Neptune from the Sun)
 at the instant
 $T=2.3 \times 10^{13}\, \text{s}\simeq 7.29\times 10^{5}\, \text{yr}$.
The profiles of distributions of the density and the velocity in the neighborhood of the outer boundary of the configuration, $r=R$,
are shown in Fig.~\ref{fig_dens_vels_col} at  different instants of time.
 Notice here that in performing numerical calculations, it is convenient to introduce new exponential function $\alpha=e^{-\bar{\alpha}}$.

By comparing the results presented in Figs.~\ref{fig_vels} and \ref{fig_dens_vels_col} with Shu's calculations shown in his
Figs.~2 and 3, it can be seen that the presence of the
chameleon scalar field with the coupling in the form \eqref{choice_f_exp_sing} does not
result in qualitative changes of the picture of the collapse.
Changes touch upon just some quantitative characteristics: starting from the task of obtaining the mass of the collapsed object
  of order a stellar mass, we have to take the external pressure 8 times smaller, and the size of the initial
 configuration $\sqrt{8}$ larger,  compared to Shu's problem.  Near the center, the process of collapse is analogous
 to the problem from Ref.~\cite{Shu:1977uc} when the total mass density
 $\rho_t \sim r^{-3/2}$ and the velocity $u \sim r^{-1/2}$
[see expressions \eqref{sols_origin_f_exp}] that provides the constant in time central accretion rate.

\section{Summary and conclusions}

In this paper, we have studied  nonrelativistic gravitating configurations
consisting of an isothermal fluid nonminimally coupled to a chameleon scalar field $\phi$,
and described by the Lagrangian \eqref{lagran_cham_star_col}.
The aim of this work was to clarify how the scalar field affects the
distribution and motion of matter in such configurations.
The crucial role here is played by the form
 of the coupling function $f=f(\phi)$. In considering these problems we restrict ourselves to the particular case
 of a massless chameleon scalar field.
 Besides allowing us to find some analytical solutions for static
 configurations, it allows us to consider the problem of the
gravitational collapse by applying the similarity method.

Namely, in Sec.~\ref{choice_f_exp} the static singular and regular solutions were found for the case of
 $f=e^{-\phi}$. It was shown  that there exist analytical singular solutions
 \eqref{sol_stat_f_exp} similar to known solutions for an isothermal sphere without a scalar field
 \cite{Chandra:1939}. The study of the regular solutions indicates that the properties of the configuration
 under consideration depend substantially on the central value of the scalar field
 $\phi_0$: (i) Depending on the sign of
$\phi_0$, the configurations will have a larger or smaller concentration of matter at the center;
(ii) For a configuration embedded in an
external medium of pressure $p_e$, the hydrostatic equilibrium has the critical value of mass \eqref{M_crit_f_exp}
which is factor of $0.44$ smaller than the critical Bonnor-Ebert sphere \cite{Ebert:1955,Bonnor:1956};
(iii) There is a critical size of the configuration, $\xi=\xi_{max}$, corresponding to the above critical mass,
whose value is determined by the value of  $\phi_0$.
In this case configurations with sizes less than
 $\xi_{max}$ may be either stable or unstable depending on the central density, and configurations with
 $\xi\geq\xi_{max}$ are always unstable.

Next, in Secs.~\ref{choice_f_pow_sing} and \ref{choice_f_pow_reg} the energy density of the system is
chosen to be a power function of the radial coordinate $\xi$ in the form \eqref{choice_fe_sing} and \eqref{choice_fe}.
For these cases,  singular and regular solutions were found. The regular solutions permit the existence of
finite configurations whose size is determined just by one free parameter -- the exponent  $\beta$
in expression \eqref{choice_fe}.
This situation differs from that in  problems
 without a chameleon scalar field when only solutions describing infinite size configurations
do exist.
 A simple stability analysis indicates that,
for the choice of $f$ in the form  \eqref{choice_fe},
 the regular, finite size solutions are unstable.

The above studies indicate that systems under consideration can be stable or unstable. In the latter case, a question arises as to
the possible motion of matter in its own gravitational field.
As an example of such motion,
in Sec.~\ref{collaps_sing} the self-similar collapse of the singular sphere with function $f$ chosen in the exponential form
\eqref{choice_f_exp_sing} was considered.
Without a scalar field, the similar problem was investigated
 by Shu in \cite{Shu:1977uc}.
 That is why we basically
 compared  our calculations
 with  Shu's results. It was shown that the qualitative behavior of the solutions
 remains the same as in the case without the scalar field:
 the process of collapse starts from the initially singular sphere
 \eqref{sing_sol_dim}
truncated at a boundary pressure $p_e$ and having the radius and the total mass given by expressions \eqref{sing_M_R}.
The collapse is initiated by an initial density perturbation  near the center
that propagates outward to the edge of the configuration in the form of the expansion wave
with the speed of sound $a$.
The presence of the scalar field gives some quantitative differences which are obviously connected with the fact that
the initial density distribution   \eqref{bound_distr_fluid} differs from Shu's distribution by the factor
$1/2$. As a result,
starting from the task of obtaining the mass of the collapsed object
  of order a stellar mass, it is necessary to set the external pressure 8 times smaller, and the size of the initial
 configuration $\sqrt{8}$ larger,  compared to Shu's problem.

As in Shu's problem, in this paper we consider
the collapse of the singular isothermal sphere  bounded by some artificial external pressure.
Without the scalar field,  inclusion of such external pressure is the only  possibility of obtaining finite size
configurations with finite masses (both for singular and regular cases).
As shown in Sec.~\ref{choice_f_pow_reg},
in the presence of the
 chameleon scalar field interacting with the fluid
 in the form  \eqref{choice_fe},
it becomes possible to find regular solutions describing more realistic finite size configurations
without the external pressure. Such configurations being initially unstable, in principle, might
be used as initial states when considering the collapse of isothermal clouds in the presence of the scalar field.

Another possibility of obtaining finite size solutions can come from a consideration of 
Eqs.~\eqref{conserv_2_cham_star_2_dmls} and \eqref{sf_cham_star_dmls} as a system which is equivalent to two interacting scalar fields
$\eta$ and $\phi$. In this case it is possible to introduce the effective Lagrangian of the system in the following form:
$$
L_{\eta, \phi}=\frac{1}{2}\partial_{\mu} \eta\, \partial^{\mu} \eta +
\frac{1}{2}\partial_{\mu} \phi\, \partial^{\mu} \phi -V(\eta, \phi),
$$
where the effective potential  $V(\eta, \phi)=-e^{-\eta} f(\phi)$. By varying this Lagrangian with respect to
 $\eta$ and $\phi$, one can obtain Eqs.~\eqref{conserv_2_cham_star_2_dmls} and \eqref{sf_cham_star_dmls}. 
 Then the problem of searching the finite size solutions
 to the field equations amounts to finding such a form
 of the coupling function $f$ that provides the required solutions, if any.

\section*{Acknowledgements}
The author is grateful to the Research Group Linkage
Programme of the Alexander von Humboldt Foundation for the
support of this research.

\appendix
\numberwithin{equation}{section}

\boldmath
\section{Derivation of the self-similar equations}
\unboldmath
\label{eqs_self_general}

In this appendix we derive equations describing self-similar motion of a perfect isothermal fluid in the presence of a
chameleon scalar field.
To do this, we start with the energy-momentum tensor
 \eqref{emt_cham_star}. Let us choose the metric in the Newtonian approximation in the form
 \begin{equation}
\label{metric_sphera_newton}
ds^2=\left(1+2\frac{\psi}{c^2}\right)c^2 dt^2-dr^2-r^2d\Omega^2,
\end{equation}
where $d\Omega^2$ is the metric on the unit 2-sphere, and $\psi$ is the Newtonian gravitational potential.
Using this metric and taking into account the expressions for the components of the four-velocity
$$
u_0=\frac{\sqrt{g_{00}}}{\sqrt{1-(v/c)^2}}\,, \quad u_1=\frac{g_{11}}{\sqrt{g_{00}}}\frac{v/c}{\sqrt{1-(v/c)^2}}\,,
$$
where $v$ is the ordinary three-dimensional velocity, one can obtain the following components of the energy-momentum tensor
\eqref{emt_cham_star} in the nonrelativistic limit:
\begin{eqnarray}
\label{comp_emt_01}
T_0^0=f\rho c^2, \quad T_1^1=-f \left(\rho v^2+p\right)-\frac{1}{2}\varphi^{\prime 2}+V(\varphi),\\
\label{comp_emt_23}
T_2^2=T_3^3=-f p+\frac{1}{2}\varphi^{\prime 2}+V(\varphi),\quad
T^1_0=-T_1^0=f \rho v c.
\end{eqnarray}
Next, using the law of conservation
$$
T^k_{i;k}=0,
$$
we get two equations: The $i=0$ component corresponding to the hydrodynamic equation of continuity
$$
\frac{1}{c}\frac{\partial T_0^0}{\partial t}+\frac{\partial T^1_0}{\partial r}+
\frac{2}{r}T^1_0=0.
$$
The $i=1$ component  (Euler's equation):
$$
\frac{1}{c}\frac{\partial T_1^0}{\partial t}+\frac{\partial T_1^1}{\partial r}+
\left(T_1^1-T_0^0\right)\frac{1}{c^2}\frac{\partial \psi}{\partial r}+\frac{2}{r}\left[T_1^1-\frac{1}{2}\left(T^2_2+T_3^3\right)\right]=0.
$$
Substituting expressions  \eqref{comp_emt_01} and \eqref{comp_emt_23} in these  equations, we have, respectively,
\begin{equation}
\label{eq_cont_1}
f\left(\frac{\partial \rho}{\partial t}+v \frac{\partial \rho}{\partial r}+\rho \frac{\partial v}{\partial r}+
\frac{2}{r}\rho v\right)+\rho\frac{\partial f}{\partial \varphi}\left(\dot{\varphi}+v\varphi^\prime\right)=0
\end{equation}
and
\begin{equation}
\label{eq_eul_1}
f\left(\frac{\partial v}{\partial t}+v \frac{\partial v}{\partial r}+\frac{1}{\rho}\frac{\partial p}{\partial r}+
\frac{\partial \psi}{\partial r}\right)+\frac{p}{\rho}\frac{\partial f}{\partial r}+
\frac{\varphi^\prime}{\rho}\left(\varphi^{\prime \prime}+\frac{2}{r}\varphi^\prime-\frac{\partial V}{\partial \varphi}\right)=0,
\end{equation}
where the prime and the dot denote  differentiation with respect to $r$ and $t$, respectively.
In the case when $f=1$ and the scalar field is absent, Eq.~\eqref{eq_cont_1} 
reduces  to the usual equation of continuity, and Eq.~\eqref{eq_eul_1}   to Euler's equation for the spherically symmetric case.

The equation for the scalar field $\varphi$ coming from the Lagrangian \eqref{lagran_cham_star_col} is
$$
\frac{1}{\sqrt{-g}}\frac{\partial}{\partial x^i}\left[\sqrt{-g}g^{ik}\frac{\partial \varphi}{\partial x^k}\right]=
-\frac{d V}{d \varphi}+L_m \frac{d f}{d \varphi}.
$$
Using this field equation with the Lagrangian for the perfect fluid $L_m=p$, and the metric
\eqref{metric_sphera_newton}, we get the following scalar field equation
\begin{equation}
\label{sf_cham_star_col}
\varphi^{\prime\prime}+\frac{2}{r}\varphi^\prime-\frac{\partial V}{\partial \varphi}=
-p\frac{\partial f}{\partial\varphi}.
\end{equation}
In obtaining this equation, we neglect the term
 $\ddot{\varphi}/c^2$, in the spirit of the nonrelativistic character of the problem.
In this approximation, Eq.~\eqref{sf_cham_star_col} is static, 
i.e. one assumes that perturbations of the scalar field propagate with infinite velocity.
Substituting Eq.~\eqref{sf_cham_star_col} in \eqref{eq_eul_1}, we get the usual Euler's equation
\begin{equation}
\label{eq_eul_2}
\frac{\partial v}{\partial t}+v \frac{\partial v}{\partial r}+\frac{1}{\rho}\frac{\partial p}{\partial r}+
\frac{\partial \psi}{\partial r}=0.
\end{equation}
Next, introducing the mass $M$ contained  within a sphere of radius $r$
\begin{equation}
\label{mass_fun}
\frac{\partial M}{\partial r}=4\pi f r^2 \rho,
\end{equation}
one can rewrite the
equation of continuity \eqref{eq_cont_1} in the following form:
$$
\frac{\partial M}{\partial t}+4\pi f r^2 \rho v=0.
$$
Substituting  expression  \eqref{mass_fun} in  the Poisson equation
$$
\frac{1}{r^2}\frac{\partial}{\partial r}\left(r^2\frac{\partial \psi}{\partial r}\right)=4\pi G f \rho,
$$
we obtain the usual equation for the Newtonian potential
$$
\frac{\partial \psi}{\partial r}=\frac{G M}{r^2}.
$$
Substituting this  in
\eqref{eq_eul_2}, we get the final form of the
set of equations describing the dynamics of the system under consideration:
\begin{eqnarray}
\label{eq_cont_2}
\frac{\partial M}{\partial t}+4\pi f r^2 \rho v&=&0,\\
\label{mass_fun_2}
\frac{\partial M}{\partial r}&=&4\pi f r^2 \rho,\\
\label{eq_eul_3}
\frac{\partial v}{\partial t}+v \frac{\partial v}{\partial r}&=&-\frac{1}{\rho}\frac{\partial p}{\partial r}-
\frac{G M}{r^2},\\
\label{sf_cham_star_col_1}
\frac{\partial^2 \varphi}{\partial r^2}+\frac{2}{r}\frac{\partial \varphi}{\partial r}-\frac{\partial V}{\partial \varphi}&=&
-p\frac{\partial f}{\partial\varphi}.
\end{eqnarray}
Here $M=M(r,t)$ is the total mass inside radius $r$ at time $t$. These equations are invariant under the time reversal
operation: $t\to -t, \quad v\to -v$
that allows us considering only the range of $0<t<\infty$. Notice here that in the absence of the nonminimal interaction,
i.e. when the coupling function $f=1$, the system becomes uncoupled:
the evolution of the fluid  does not affect the distribution  of the scalar field. This situation differs
from the case of the presence of strong gravitational fields considered in Ref.~\cite{arXiv:1106.1267}
when even at  $f=1$
equations describing  static configurations are not independent, but are coupled via gravity.

In the case of the polytropic equation of state
 with polytropic exponent $\gamma$
\begin{equation}
\label{eqs_cham_star_col}
p=K \rho^\gamma,
\end{equation}
where $K$  is some positive arbitrary constant,
Eqs.~\eqref{eq_cont_2}-\eqref{sf_cham_star_col_1} permit the introduction of similarity variables given by \cite{Suto:1988}:
\begin{align}
\label{self_variab}
\begin{split}
x&=\frac{r}{\sqrt{k}t^n},\quad v(r,t)=\sqrt{k}t^{n-1}u(x), \quad \rho(r,t)=\frac{\alpha(x)}{4\pi G t^2}, \\
p(r,t)&=\frac{k}{4\pi G}\,t^{2n-4}[\alpha(x)]^\gamma, \quad M(r,t)=\frac{k^{3/2} t^{3n-2}}{(3n-2)G}\,m(x), \quad
\varphi(r,t)= \frac{K}{\sqrt{4\pi G}}\,\phi(x),
\end{split}
\end{align}
where $k$ is some dimensional constant, and $n$ -- a dimensionless constant. Rewriting Eqs.~\eqref{eq_cont_2} and
\eqref{mass_fun_2} in terms of the above dimensionless similarity variables, we get an algebraic expression for $m(x)$ in terms of
$\alpha(x)$ and $u(x)$
\begin{equation}
\label{mass_fun_algeb}
m(x)=f x^2 \alpha(n x-u).
\end{equation}
Next, Eq.~\eqref{eq_eul_3} gives
\begin{equation}
\label{eq_eul_4}
\gamma\alpha^{\gamma-2}\frac{d\alpha}{dx}-(n x-u)\frac{du}{dx}=-(n-1)\,u-f\frac{n x-u}{3n-2}\,\alpha.
\end{equation}
Differentiating now Eq.~\eqref{mass_fun_algeb} with respect to  $x$
and taking into account  \eqref{mass_fun_2}, we get
\begin{equation}
\label{eq_mixed}
(n x-u)\frac{d\alpha}{dx}-\alpha\frac{d u}{dx}=-2\frac{x-u}{x}\,\alpha-\alpha(n x-u)\frac{d \ln{f}}{d\phi}\frac{d\phi}{d x}.
\end{equation}
Next, Eq.~\eqref{sf_cham_star_col_1} takes the form
\begin{equation}
\label{sf_cham_star_col_2}
\frac{d^{2}\phi}{d x^2}+\frac{2}{x}\frac{d\phi}{d x}=\frac{4\pi G k}{K^2} t^{2n}\frac{d V}{d\phi}-
\frac{k^2}{K^2}t^{4(n-1)}\alpha^{\gamma}\frac{d f}{d \phi}.
\end{equation}

It can be seen from Eqs.~\eqref{eqs_cham_star_col} and \eqref{self_variab}
that the above similarity variables permit the following relation:
\begin{equation}
\label{k_restr}
K=k(4\pi G)^{\gamma-1}t^{2(n+\gamma-2)}.
\end{equation}
Taking this expression into account, one can see from
\eqref{sf_cham_star_col_2} that it is impossible to exclude the time variable
 $t$ in front of $d V/d\phi$ and
 $d f/d\phi$ simultaneously (as required by the self-similar character of the motion under consideration).
That is why, bearing in mind that we look for solutions with the function  $f\neq 1$, we
will consider  a massless scalar field with  $V(\phi)=0$ only.
In this case the last term on the right-hand side of Eq.~\eqref{sf_cham_star_col_2} is proportional to
 $t^{4(1-\gamma)}$. Hence the exponent is equal to zero when $\gamma=1$.
Taking this into account and requiring  that the parameter
 $K$ from \eqref{k_restr} be independent of time,
i.e. that $(n+\gamma-2)=0$, it is necessary to take $n=1$. Then we finally have from
 \eqref{sf_cham_star_col_2}
\begin{equation}
\label{sf_cham_star_col_3}
\frac{d^2\phi}{d x^2}+\frac{2}{x}\frac{d\phi}{d x}=
-\alpha\frac{d f}{d \phi}.
\end{equation}
Letting $\gamma=n=1$, Eqs.~\eqref{eq_eul_4} and \eqref{eq_mixed}, in turn,   can be rewritten in the following form:
\begin{eqnarray}
\label{eq_fin_1}
&&\frac{d\alpha}{d x}=\frac{\alpha}{(x-u)^2-1}\left[f\alpha-2\frac{x-u}{x}-(x-u)\frac{d\ln{f}}{d\phi}\frac{d\phi}{dx}\right](x-u),\\
\label{eq_fin_2}
&&\frac{d u}{d x}=\frac{1}{(x-u)^2-1}\left[f\alpha(x-u)-\frac{2}{x}-\frac{d\ln{f}}{d\phi}\frac{d\phi}{dx}\right](x-u),\\
\label{eq_fin_3}
&&\frac{d^2\phi}{d x^2}+\frac{2}{x}\frac{d\phi}{d x}=
-\alpha\frac{d f}{d \phi}.
\end{eqnarray}
When the scalar field is absent, the system
 \eqref{eq_fin_1}-\eqref{eq_fin_3} describes  the known problem of the self-similar motion of an isothermal fluid
 considered, e.g. in Refs.~\cite{Larson:1969mx,Penston:1969yy,Shu:1977uc,Hunter:1977}.

\end{document}